\documentclass[notitlepage,a4paper,aps,prd,onecolumn,superscriptaddress,nofootinbib,groupedaddress]{revtex4}
\usepackage{amsmath}
\usepackage{amsfonts,color}
\usepackage{amsmath}
\usepackage{amsthm}
\usepackage{pdfpages}
\usepackage{amsfonts,color}
\usepackage{amssymb,float}

\usepackage{amsmath}

\DeclareMathOperator{\arccosh}{arccosh}

\usepackage[utf8]{inputenc}
\allowdisplaybreaks
\usepackage{color}
\usepackage{hyperref}
\hypersetup{
    colorlinks=true,
    linkcolor=blue,
    filecolor=magenta,      
   citecolor=blue
}

\usepackage{enumitem}

\usepackage{tikz}
\usetikzlibrary{shapes.geometric}
\usetikzlibrary{arrows.meta,arrows}

\begin{document}
\title{Rotating Kerr-Newman space-times in Metric-Affine Gravity}

\author{Sebastian Bahamonde}
\email{sbahamonde@ut.ee, sebastian.beltran.14@ucl.ac.uk}
\affiliation{Laboratory of Theoretical Physics, Institute of Physics, University of Tartu, W. Ostwaldi 1, 50411 Tartu, Estonia.}

\author{Jorge Gigante Valcarcel}
\email{jorge.gigante.valcarcel@ut.ee}
\affiliation{Laboratory of Theoretical Physics, Institute of Physics, University of Tartu, W. Ostwaldi 1, 50411 Tartu, Estonia.}

\begin{abstract}

We present new rotating vacuum configurations endowed with both dynamical torsion and nonmetricity fields in the framework of Metric-Affine gauge theory of gravity. For this task, we consider scalar-flat Weyl-Cartan geometries and obtain an axisymmetric Kerr-Newman solution in the decoupling limit between the orbital and the spin angular momentum. The corresponding Kerr–Newman–de Sitter solution is also compatible with a cosmological constant and additional electromagnetic fields.

\end{abstract}

\maketitle

\section{Introduction}
The recent discoveries of gravitational waves and the first evidence of the shadow observed from a supermassive black hole in M87 have opened a new era for black hole physics~\cite{Akiyama:2019cqa,Abbott:2016blz,TheLIGOScientific:2017qsa,LIGOScientific:2021qlt}. General Relativity (GR) predicts that the simplest rotating black hole solution can be described by three main parameters: its mass, angular momentum and electric charge. Since at astrophysical scales it is expected that this electric charge is almost negligible, the most common way to describe a rotating black hole solution is by the Kerr space-time, which is fully determined by its mass and angular momentum \cite{Kerr:1963ud}. A vast number of  studies concerning astrophysical observations are based on this solution since it can describe different observations in a good agreement~\cite{Broderick:2013rlq,Cardoso:2016ryw,Isi:2019aib,Johannsen:2015hib}. In fact, even though the assumption of static and spherically symmetric space-times can provide us a good understanding on the physics of compact objects, the search of stationary rotating solutions turns out to be essential for a full phenomenological assessment in terms of a realistic configuration \cite{Visser:2007fj,Bahamonde:2020snl}.

There are different evidences from observations to theory pointing out that GR does not constitute the final theory of gravity. Some indications are for example: the growing tension in observations of the value of the expansion of the universe determined by the Hubble constant $H_0$, the problem of dark energy and dark matter, or its inconsistency at the quantum level~\cite{Planck:2018vyg,Wong:2019kwg,DiValentino:2021izs,Peebles:2002gy,Navarro:1995iw}. This leads to consider further extensions and modifications of GR (see~\cite{Nojiri:2017ncd,Joyce:2014kja,CANTATA:2021ktz} for recent reviews on this topic). In particular, the establishment of the gauge principles in the realm of post-Riemannian geometry displays the torsion and nonmetricity tensors sourced by the hypermomentum of matter as new properties of the gravitational field, which gives rise to the formulation of Metric-Affine Gauge theory of gravity (MAG)~\cite{Hehl:1994ue,Blagojevic:2013xpa,Cabral:2020fax}.

The study of axial symmetry in MAG is not so vast due to its complexity. Exact solutions have been found in certain models assuming the so-called triplet ansatz and other prolongation techniques \cite{Vlachynsky:1996zh,Hehl:1999sb,Baekler:2006de}. Recently, the authors of this work followed a different approach based on consistency constraints and found a new spherically symmetric solution with independent dynamical torsion and nonmetricity fields beyond the triplet ansatz~\cite{Bahamonde:2020fnq}. This result generalises the previous findings obtained in the Poincar\'e gauge framework of MAG, where it was shown that the vacuum structure of GR is modified in the presence of a Coulomb-like dynamical torsion~\cite{Cembranos:2016gdt,Cembranos:2017pcs} (see~\cite{Blagojevic:2021pqp} for a consistent thermodynamic analysis and~\cite{Obukhov:2020hlp} for a natural extension including parity odd terms). Thereby, the aim of this work is to study axial symmetry and to obtain new rotating black hole solutions in the metric-affine geometry described by MAG.

This work is organised as follows. In Sec.~\ref{sec:action} we briefly describe the general features of metric-affine geometry and the model with higher order corrections recently proposed to display the independent dynamical behaviour of the torsion and nonmetricity fields. Then, in Sec.~\ref{subsec:axialA} we set the invariance conditions of stationary and axisymmetric post-Riemannian geometries, and perform a decomposition into propagating and non-propagating modes, in order to present a consistent method for solving the field equations of the model and simplifying the problem. This allows us to obtain in Sec.~\ref{subsec:axialB} a new Kerr-Newman (KN) black hole solution with torsion and nonmetricity, under the assumption that the coupling between the orbital and the spin angular momentum of the solution is small. Finally, we conclude our main results in Sec.~\ref{sec:conclusions}. We work in the natural units $c=G=1$ with the metric signature $(+,-,-,-)$. Geometric quantities with a tilde on top denote that they are computed with respect to a general affine connection whereas quantities without a tilde denote that they are computed with respect to the Levi-Civita connection. On the other hand, Latin and Greek indices refer to anholonomic and coordinate basis, respectively.

\section{Metric-Affine Gravity with dynamical torsion and nonmetricity}\label{sec:action}

The formulation of gravity within an affinely connected metric space-time allows the post-Riemannian degrees of freedom present in a general affine connection to be considered as additional properties of the gravitational interaction. Specifically, such a general affine connection provides not only curvature changes in the geometry of the space-time but also torsion and nonmetricity deformations defined as follows:
\begin{align}
    T^{\lambda}\,_{\mu \nu}&=2\tilde{\Gamma}^{\lambda}\,_{[\mu \nu]}\,,\\
    Q_{\lambda \mu \nu}&=\tilde{\nabla}_{\lambda}g_{\mu \nu}\,.
\end{align}

Then, the covariant derivative of an arbitrary vector $v^{\lambda}$ can be split into a Riemannian contribution and a distortion tensor
\begin{equation}
\tilde{\nabla}_{\mu}v^{\lambda}=\nabla_{\mu}v^{\lambda}+N^{\lambda}\,_{\rho\mu}v^{\rho}\,,
\end{equation}
which encodes the torsion and nonmetricity pieces as the sum of the so-called contortion and disformation tensors
\begin{equation}
N^{\lambda}\,_{\rho\mu}=K^{\lambda}\,_{\rho\mu}+L^{\lambda}\,_{\rho\mu}\,,
\end{equation}
with
\begin{align}
    K^{\lambda}\,_{\rho\mu}&=\frac{1}{2}(T^{\lambda}\,_{\rho\mu}-T_{\rho}\,^{\lambda}\,_{\mu}-T_{\mu}\,^{\lambda}\,_{\rho})\,,
\end{align}
\begin{align}
    L^{\lambda}\,_{\rho\mu}&=\frac{1}{2}(Q^{\lambda}\,_{\rho\mu}-Q_{\rho}\,^{\lambda}\,_{\mu}-Q_{\mu}\,^{\lambda}\,_{\rho})\,.
\end{align}
The corrections modify not only the definition of the covariant derivative but also its commutation relations, which provide the notion of the intrinsic curvature in terms of the affine connection
\begin{equation}
[\tilde{\nabla}_{\mu},\tilde{\nabla}_{\nu}]\,v^{\lambda}=\tilde{R}^{\lambda}\,_{\rho \mu \nu}\,v^{\rho}+T^{\rho}\,_{\mu \nu}\,\tilde{\nabla}_{\rho}v^{\lambda}\,,
\end{equation}
where
\begin{equation}\label{totalcurvature}
\tilde{R}^{\lambda}\,_{\rho \mu \nu}=\partial_{\mu}\tilde{\Gamma}^{\lambda}\,_{\rho \nu}-\partial_{\nu}\tilde{\Gamma}^{\lambda}\,_{\rho \mu}+\tilde{\Gamma}^{\lambda}\,_{\sigma \mu}\tilde{\Gamma}^{\sigma}\,_{\rho \nu}-\tilde{\Gamma}^{\lambda}\,_{\sigma \nu}\tilde{\Gamma}^{\sigma}\,_{\rho \mu}\,.
\end{equation}

The presence of torsion and nonmetricity in the space-time requires gauging the external degrees of freedom consisting of translations, rotations, dilations and shears by means of the affine group $A(4,R) = R^{4} \rtimes GL(4,R)$ \cite{Hehl:1994ue,Blagojevic:2013xpa,Cabral:2020fax}. Indeed, a gauge approach to gravity is successfully formulated when the unitary irreducible representations of particles are linked to the geometry of the space-time. In particular, the absence of spinor representations for general coordinate transformations demands an anholonomic spin connection as the fundamental quantity for the introduction of
spinor fields into the geometry of the space-time, which in the framework of MAG is associated with the general metalinear group $\overline{GL}(4,R)$ \cite{Neeman:1977iup,Neeman:1987pzd}.

Thereby, the nontrivial structure of the group $A(4,R)$ allows the dynamical behaviour of torsion and nonmetricity to be performed in terms of a large number of geometric invariants in the gravitational action. Specifically, in the case of Weyl-Cartan geometry the traceless parts of the nonmetricity tensor vanish and this tensor is fully ascribed to the so-called Weyl vector
\begin{equation}
W_{\mu}=\frac{1}{4}\,Q_{\mu\nu}\,^{\nu}
\,.
\end{equation}
Then, the gauge group $A(4,R)$ is reduced to the inhomogeneous Weyl group $IW(1,3)$ and it is possible to define the following model based on scalar-flat Weyl-Cartan geometries to display such a behaviour conforming a black hole solution within a spherically symmetric space-time \cite{Bahamonde:2020fnq}:
\begin{eqnarray}\label{LagrangianIrreducible}
S &=& \frac{1}{64\pi}\int d^4x \sqrt{-g}
\left.\Bigl[
-\,4R-6d_{1}\tilde{R}_{\lambda\left[\rho\mu\nu\right]}\tilde{R}^{\lambda\left[\rho\mu\nu\right]}-9d_{1}\tilde{R}_{\lambda\left[\rho\mu\nu\right]}\tilde{R}^{\mu\left[\lambda\nu\rho\right]}+8\,d_{1}\tilde{R}_{\left[\mu\nu\right]}\tilde{R}^{\left[\mu\nu\right]}
\Bigr.
\right.
\nonumber\\
& &
\left.
\Bigl.
\;\;\;\;\;\;\;\;\;\;\;\;\;\;\;\;\;\;\;\;\;\;\;\;\;\;\;\;\;+\,\frac{1}{8}\left(32e_{1}+13d_{1}\right)\tilde{R}^{\lambda}\,_{\lambda\mu\nu}\tilde{R}^{\rho}\,_{\rho}\,^{\mu\nu}-7d_{1}\tilde{R}_{\left[\mu\nu\right]}\tilde{R}^{\lambda}\,_{\lambda}\,^{\mu\nu}\Bigr]\right.\,.
\end{eqnarray}

Such a black hole solution must satisfy the following field equations, derived from the variations of the Expression~(\ref{LagrangianIrreducible}) with respect to the vierbein field and the anholonomic connection:
\begin{eqnarray}\label{field_eq1}
0 &=& 2\,G_{\mu}\,^{\nu}+16\pi\tilde{\mathcal{L}}\,\delta_{\mu}\,^{\nu}+2d_{1}\left(\tilde{R}^{\nu}\,_{\lambda\rho\mu}\tilde{R}^{\left[\lambda\rho\right]}+\tilde{R}_{\lambda}\,^{\nu}\,_{\mu\rho}\hat{R}^{\left[\lambda\rho\right]}+\tilde{R}_{\lambda\mu}\tilde{R}^{\left[\nu\lambda\right]}+\hat{R}_{\lambda\mu}\hat{R}^{\left[\nu\lambda\right]}\right)
\nonumber\\
&&+\frac{d_1}{2}\tilde{R}_{\lambda\rho\sigma\mu}
\Bigl[
\left(4\tilde{R}^{\left[\nu\sigma\right]\lambda\rho}-2\tilde{R}^{\left[\lambda\rho\right]\nu\sigma}-\tilde{R}^{\left[\lambda\nu\right]\rho\sigma}-\tilde{R}^{\left[\rho\sigma\right]\lambda\nu}-\tilde{R}^{\left[\rho\nu\right]\sigma\lambda}-\tilde{R}^{\left[\sigma\lambda\right]\rho\nu}\right)
\Bigr.\,
\nonumber\\
\Bigl.
&&-\,4\left(\tilde{R}^{\left(\rho\nu\right)\lambda\sigma}-\tilde{R}^{\left(\lambda\sigma\right)\rho\nu}+\tilde{R}^{\left(\lambda\nu\right)\rho\sigma}-\tilde{R}^{\left(\rho\sigma\right)\lambda\nu}\right)
\Bigr]-4e_{1}\tilde{R}^{\lambda}\,_{\lambda\sigma\mu}\tilde{R}^{\rho}\,_{\rho}\,^{\sigma\nu}\,,
\end{eqnarray}
\begin{eqnarray}\label{field_eq2}
0&=&2d_{1}
\Bigl[
\nabla_{\rho}\left(g^{\mu\nu}\tilde{R}^{\left[\lambda\rho\right]}-g^{\lambda\nu}\hat{R}^{\left[\mu\rho\right]}+g^{\lambda\rho}\hat{R}^{\left[\mu\nu\right]}-g^{\mu\rho}\tilde{R}^{\left[\lambda\nu\right]}\right)+N^{\rho\mu}\,_{\rho}\tilde{R}^{\left[\lambda\nu\right]}-N^{\rho\lambda}\,_{\rho}\hat{R}^{\left[\mu\nu\right]}
\Bigr.
\nonumber\\
\Bigl.
&&+N^{\nu\lambda}\,_{\rho}\hat{R}^{\left[\mu\rho\right]}-N^{\nu\mu}\,_{\rho}\tilde{R}^{\left[\lambda\rho\right]}+N^{\mu}\,_{\rho}\,^{\lambda}\hat{R}^{\left[\rho\nu\right]}-N^{\lambda}\,_{\rho}\,^{\mu}\tilde{R}^{\left[\rho\nu\right]}
\Bigr]
\nonumber\\
&&+\frac{d_1}{2}\Bigl(\nabla_{\rho}+W_{\rho}\Bigr)
\Bigl[
\left(4\tilde{R}^{\left[\rho\nu\right]\lambda\mu}-2\tilde{R}^{\left[\lambda\mu\right]\rho\nu}-\tilde{R}^{\left[\mu\nu\right]\lambda\rho}+\tilde{R}^{\left[\lambda\nu\right]\mu\rho}-\tilde{R}^{\left[\lambda\rho\right]\mu\nu}+\tilde{R}^{\left[\mu\rho\right]\lambda\nu}\right)
\Bigr.
\nonumber\\
\Bigl.
&&-\,4\left(\tilde{R}^{\left(\mu\nu\right)\lambda\rho}-\tilde{R}^{\left(\lambda\rho\right)\mu\nu}+\tilde{R}^{\left(\nu\lambda\right)\mu\rho}-\tilde{R}^{\left(\rho\mu\right)\lambda\nu}\right)
\Bigr]
-4e_{1}g^{\lambda \mu}\nabla_{\rho}\tilde{R}_{\sigma}\,^{\sigma\rho\nu}
\nonumber\\
&&+\frac{d_1}{2}N^{\lambda}\,_{\sigma\rho}
\Bigl[
\left(4\tilde{R}^{\left[\rho\nu\right]\sigma\mu}-2\tilde{R}^{\left[\sigma\mu\right]\rho\nu}-\tilde{R}^{\left[\mu\nu\right]\sigma\rho}+\tilde{R}^{\left[\sigma\nu\right]\mu\rho}-\tilde{R}^{\left[\sigma\rho\right]\mu\nu}+\tilde{R}^{\left[\mu\rho\right]\sigma\nu}\right)
\Bigr.
\nonumber\\
\Bigl.
&&-\,4\left(\tilde{R}^{\left(\mu\nu\right)\sigma\rho}-\tilde{R}^{\left(\sigma\rho\right)\mu\nu}+\tilde{R}^{\left(\nu\sigma\right)\mu\rho}-\tilde{R}^{\left(\rho\mu\right)\sigma\nu}\right)+4g^{\mu\nu}\tilde{R}^{\left[\sigma\rho\right]}
\Bigr]
\nonumber\\
&&+\frac{d_1}{2}N^{\mu}\,_{\sigma\rho}
\Bigl[
\left(4\tilde{R}^{\left[\rho\nu\right]\lambda\sigma}-2\tilde{R}^{\left[\lambda\sigma\right]\rho\nu}-\tilde{R}^{\left[\sigma\nu\right]\lambda\rho}+\tilde{R}^{\left[\lambda\nu\right]\sigma\rho}-\tilde{R}^{\left[\lambda\rho\right]\sigma\nu}+\tilde{R}^{\left[\sigma\rho\right]\lambda\nu}\right)
\Bigr.
\nonumber\\
\Bigl.
&&-\,4\left(\tilde{R}^{\left(\sigma\nu\right)\lambda\rho}-\tilde{R}^{\left(\lambda\rho\right)\sigma\nu}+\tilde{R}^{\left(\nu\lambda\right)\sigma\rho}-\tilde{R}^{\left(\sigma\rho\right)\lambda\nu}\right)-4g^{\lambda\nu}\hat{R}^{\left[\sigma\rho\right]}
\Bigr]\,,
\end{eqnarray}
where $\tilde{\mathcal{L}}$ represents the quadratic order of the Lagrangian density. It is worthwhile to stress that the algebraic structure of the inhomogeneous Weyl group generally provides a more complex dynamics for the geometric degrees of freedom of the space-time, in comparison with the one provided by the compact classical groups of internal symmetries \cite{Kopczynski:1988jq}. In particular, in the present MAG model the dynamical behaviour of the torsion and nonmetricity fields is described in the gravitational action~(\ref{LagrangianIrreducible}) by the field strength tensors
\begin{equation}\label{curvbianchi}
\tilde{R}^{\lambda}\,_{[\mu \nu \rho]}=\tilde{\nabla}_{[\mu}T^{\lambda}\,_{\rho\nu]}+T^{\sigma}\,_{[\mu\rho}\,T^{\lambda}\,_{\nu] \sigma}\,,\end{equation}
\begin{equation}\label{riccibianchi}
\tilde{R}_{[\mu \nu]}=\frac{1}{2}\tilde{R}^\lambda\,_{\lambda\mu\nu}+\tilde{\nabla}_{[\mu}T^{\lambda}\,_{\nu]\lambda}+\frac{1}{2}\tilde{\nabla}_{\lambda}T^{\lambda}\,_{\mu\nu}-\frac{1}{2}T^{\lambda}\,_{\rho\lambda}T^{\rho}\,_{\mu\nu}\,,
\end{equation}
\begin{equation}\label{nonmetricitybianchi}
\tilde{R}^{\lambda}\,_{\lambda\mu\nu}=4\nabla_{[\nu}W_{\mu]}\,,
\end{equation}
which describe the deviation of the Bianchi identities of GR provided by Weyl-Cartan geometry. In any case, the explicit relation between the antisymmetric part of the Ricci tensor and the homothetic component of the curvature tensor present in Expression~(\ref{riccibianchi}) clearly indicates that the former contains propagating terms of both torsion and nonmetricity tensors, which can also be implicitly present in Expression~(\ref{curvbianchi}) by means of the torsion tensor itself (i.e. in terms of dynamical constraints involving certain irreducible modes of torsion that suppress their propagating behaviour). Accordingly, depending on their dynamical contribution to the gauge field strength tensors, the irreducible parts of the torsion tensor, namely the trace vector $T_{\mu}=T^{\nu}\,_{\mu\nu}$, the axial vector $S_{\mu}=\varepsilon_{\mu\lambda\rho\nu}T^{\lambda\rho\nu}$, and the traceless and pseudotraceless tensor $t_{\lambda\mu\nu}=T_{\lambda\mu\nu}-\frac{1}{3}\left(g_{\lambda\nu}T_{\mu}-g_{\lambda\mu}T_{\nu}\right)-\frac{1}{6}\,\varepsilon_{\lambda\rho\mu\nu}S^{\rho}$, can be split into propagating and non-propagating modes, respectively denoted by a bar and a circle on top:
\begin{eqnarray}\label{decomposition2}
T_{\mu}&=&\mathring{T}_{\mu}+\bar{T}_{\mu}\,,
\\
S_{\mu}&=&\mathring{S}_{\mu}+\bar{S}_{\mu}\,,
\\
t_{\lambda\mu\nu}&=&\mathring{t}_{\lambda\mu\nu}+\bar{t}_{\lambda\mu\nu}\,.
\end{eqnarray}
The latter can be characterised by pure gauge configurations with vanishing or degenerate field strength tensors, which allows us to express the nontrivial antisymmetrized part of the curvature tensor as a general sum of a degenerate homothetic curvature and an independent dynamical component:
\begin{equation}\label{curvaturedecomposition}
\tilde{R}^{\lambda}\,_{[\mu \nu \rho]}=\mathring{R}^{\lambda}\,_{[\mu \nu \rho]}+\bar{R}^{\lambda}\,_{[\mu \nu \rho]}\,,
\end{equation}
with $\mathring{R}^{\lambda}\,_{[\mu \nu \rho]}=\alpha\tilde{R}^{\sigma}\,_{\sigma[\mu\nu}\delta_{\rho]}\,^{\lambda}$ and $\alpha$ is a constant. The consideration of this decomposition for the present model shall allow us to clarify the construction of axisymmetric solutions, as is shown in the following section.

\section{Axial symmetry and rotating space-times in Metric-Affine Gravity}\label{sec:axial}

\subsection{Invariance conditions and structure of the solution}\label{subsec:axialA}

The geometry of stationary and axisymmetric space-times is characterised by two Killing vectors $\partial_{t}$ and $\partial_{\varphi}$, which generate time translations and rotations around a symmetry axis, respectively. Accordingly, the latter defines a regular two-dimensional timelike surface of fixed points where it vanishes and provides a metric structure which is invariant under the action of the rotation group SO(2) \cite{Stephani:2003tm,Ortin:2015hya}. Given the fact that these two Killing vectors are not orthogonal to each other, it is possible to write the most general stationary and axisymmetric space-time in a specific gauge with only one nonvanishing off-diagonal component $g_{t\varphi}$ \cite{hartle1967variational}:
\begin{equation}\label{axi_line}
    ds^2=\Psi_1(r,\vartheta)\,dt^2-\frac{dr^2}{\Psi_2(r,\vartheta)}-r^2\Psi_3(r,\vartheta)\Big[ d\vartheta^2+\sin^2\vartheta(d\varphi-\Psi_4(r,\vartheta)dt)^2\Big]\,,
\end{equation}
where $(t,r,\vartheta,\varphi)$ denote spherical coordinates and $\{\Psi_{i}\}_{i=1}^{4}$ are four arbitrary functions depending on $r$ and $\vartheta$. For the present MAG model, the scalar-flatness condition fulfilled by the gravitational action (\ref{LagrangianIrreducible}) means that the Riemannian scalar curvature vanishes, which represents a strong geometrical constraint involving the metric tensor alone and therefore restricts the form of Expression~(\ref{axi_line}). In particular, the KN space-time describes a geometry compatible with a charged rotating mass fulfilling the constraint $R=0$, which constitutes the minimal extension that one can consider for the study of rotation and axial symmetry in the present model. Thereby, by assuming Boyer-Lindquist coordinates $(\tilde{t},\tilde{r},\tilde{\vartheta},\tilde{\varphi})$ \footnote{Note that we assume Boyer-Lindquist coordinates in the following, and accordingly we omit the use of tilde in our coordinates notation.}
\begin{eqnarray}
t=\tilde{t}\,,\quad r=\sqrt{\tilde{r}^2+a^2\sin^2\tilde{\vartheta}}\,,\quad \cos\vartheta=\frac{\tilde{r}}{r}\cos\tilde{\vartheta}\,,\quad \varphi=\tilde{\varphi}\,,
\end{eqnarray}
the line element~(\ref{axi_line}) and the tetrad fields in the orthonormal gauge can be set as follows:
\begin{align}\label{metric}
    ds^2&=\Psi(r,\vartheta)\,dt^2-\frac{r^2+a^2\cos^{2}\vartheta}{\left(r^2+a^2\cos^{2}\vartheta\right)\Psi(r,\vartheta)+a^2\sin^{2}\vartheta}\,dr^2-\left(r^2+a^2\cos^{2}\vartheta\right)d\vartheta^2\nonumber\\
    &-\sin^{2}\vartheta\left[r^2+a^2+a^2\left(1-\Psi(r,\vartheta)\right)\sin^{2}\vartheta\right]d\varphi^2+2a\left(1-\Psi(r,\vartheta)\right)\sin^{2}\vartheta\,dtd\varphi\,,
\end{align}
\begin{align}\label{orthoframe}
    e^a{}_\mu = 
    \left(
    \begin{array}{cccc}
    1/\sqrt{g_{rr}(r,\vartheta)} & 0 & 0 & -\,a\sin^{2}\vartheta/\sqrt{g_{rr}(r,\vartheta)}\\
    0 & \sqrt{g_{rr}(r,\vartheta)} & 0 & 0\\
    0 & 0 & \sqrt{r^2+a^2\cos^{2}\vartheta} & 0\\
    -\,a\sin\vartheta/\sqrt{r^2+a^2\cos^{2}\vartheta} & 0 & 0 & \left(r^2+a^2\right)\sin\vartheta/\sqrt{r^2+a^2\cos^{2}\vartheta}\\
    \end{array}
    \right)\,,
\end{align}
where $\Psi(r,\vartheta)=1-(2mr-k)/(r^2+a^2\cos^{2}\vartheta)$ depends on the mass $m$, the orbital angular momentum (per unit mass) $a$ and a constant parameter $k$, which in the Einstein-Maxwell model of GR can be related to electric and magnetic charges. In the realm of metric-affine geometry with independent metric, torsion and nonmetricity fields, the invariance defined by the Killing vectors can be in any case extended to post-Riemannian quantities, in such a way that the corresponding curvature tensor fulfills the same symmetries. Then, the conditions provided by the vanishment of the Lie derivatives of the torsion and nonmetricity tensors under the time and axial Killing vectors of stationary and axisymmetric space-times are realised by
\begin{eqnarray}
T_{\lambda\mu\nu}&=&T_{\lambda\mu\nu}(r,\vartheta)\label{ICtorsion}\,,\\
W_{\mu}&=&W_{\mu}(r,\vartheta)\label{ICnonmetricity}\,.
\end{eqnarray}

Therefore, the number of degrees of freedom involved for the resolution of the field equations (\ref{field_eq1}) and (\ref{field_eq2}) is twenty four for the torsion sector and four for the Weyl vector, which means that additional assumptions must be considered in order to obtain an axisymmetric configuration. In this regard, the aforementioned decomposition~(\ref{decomposition2}) into propagating and non-propagating modes can be considered to construct such solutions and simplify the problem, since the anholonomic connection must obey the same type of structure, in virtue of its correspondence with the affine connection:
\begin{equation}
    \omega^{a b}\,_{\mu}=\sum_{i=1}^{3}\omega_{i}^{a b}\,_{\mu}\,,
\end{equation}
where
\begin{align}
    \omega_{1}^{a b}\,_{\mu}&=e^{a}\,_{\lambda}\,e^{b \rho}\left(\Gamma^{\lambda}\,_{\rho \mu}+\mathring{K}_{1}^{\lambda}\,_{\rho\mu}\right)+e^{a}\,_{\lambda}\,\partial_{\mu}\,e^{b \lambda}\,,\\
    \omega_{2}^{a b}\,_{\mu}&=e^{a}\,_{\lambda}\,e^{b \rho}\left(\mathring{K}_{2}^{\lambda}\,_{\rho\mu}+L^{\lambda}\,_{\rho\mu}\right)\,,\\
    \omega_{3}^{a b}\,_{\mu}&=e^{a}\,_{\lambda}\,e^{b \rho}\bar{K}^{\lambda}\,_{\rho\mu}\,.
\end{align}
In particular, concerning the non-propagating part of the post-Riemannian degrees of freedom, the first component $\omega_{1}$ of the anholonomic connection is set to generate vanishing field strength tensors, whereas $\omega_{2}$ provides a homothetic field strength tensor associated with the scale part of the Weyl group and accordingly the antisymmetric contribution of the Weyl vector to the spin connection must be compensated by an auxiliary vector mode of torsion, which gives rise to a degenerate homothetic component in the antisymmetrized part of the curvature tensor. Additionally, the component $\omega_{3}$ is aimed to describe the independent parts of the field strength tensors yielded by the torsion field.

In the first case, the condition for a vanishing antisymmetrized part of the curvature tensor can then be written as
\begin{equation}
    \tilde{\nabla}_{[\mu}\mathring{T}_{1}^{\lambda}\,_{\rho\nu]}+\mathring{T}_{1}^{\sigma}\,_{[\mu\rho}\,\mathring{T}_{1}^{\lambda}\,_{\nu]\sigma}=0\,,
\end{equation}
but due to the high complexity of axial symmetry in metric-affine geometry it is not possible to obtain directly a general solution for the components of the torsion tensor constrained by this equation and one has to consider additional assumptions, which allow a solution to be found. Following these lines, we assume the premise that the anholonomic connection $\omega_{1}$ related to these torsion components takes Minkowski values in a certain local frame and does not depend on any of the physical parameters of the solution (namely the mass, orbital angular momentum, and spin and dilation charges), in such a way that its contribution to the field strength tensors is identically zero and the MAG model coincides with GR in the absence of propagating modes. The construction requires then the existence of a local Lorentz transformation fitting this property in the spin connection, which for the same type of scalar-flat models studied with spherically symmetric conditions was shown to be the case \cite{Cembranos:2016gdt,Cembranos:2017pcs,Bahamonde:2020fnq}, whereas for rotating axisymmetric space-times involves a further generalisation, as shown below.

In the first place, it is straightforward to check that the Minkowski space-time expressed in Boyer-Lindquist coordinates is recovered by the line element~(\ref{metric}) and the orthonormal frame~(\ref{orthoframe}) if $\Psi(r,\vartheta)=1$, in such a way that the underlying metric structure is flat. By contrast, from a mathematical point of view the torsion tensor is not subject to a unique specific form, but constrained by the vanishment of its field strength tensors (\ref{curvbianchi}) and (\ref{riccibianchi}), as well as by the boundary conditions presented in the spherically symmetric case \cite{Cembranos:2016gdt,Cembranos:2017pcs,Bahamonde:2020fnq}. In this sense, the additional assumption of a local frame where the corresponding non-propagating part $\hat{\omega}_{1}$ of the spin connection, denoted by hat to indicate that is evaluated in the rotated frame, does not depend on any of the physical parameters of the solution allows us to set these boundary conditions as its Minkowski values in the Lorentz Lie algebra, namely
\begin{eqnarray}\label{MinkowskiSP}
\hat{\omega}_{1}=
\frac{1}{2}\left(J_{\hat{1}\hat{2}}-J_{\hat{0}\hat{2}}\right)\,d\vartheta +\frac{1}{2}\sin\vartheta \left[\left(J_{\hat{1}\hat{3}}-J_{\hat{0}\hat{3}}\right)+2\cot\vartheta J_{\hat{2}\hat{3}} \right]\,d\varphi\,,
\end{eqnarray}
where $J_{a b}$ are the generators of local Lorentz rotations. Then, the transition to a KN type of geometry can be performed by considering the existence of a more general rotated basis $\vartheta^{a}=\Lambda^{a}\,_{b}\,e^{b}$ depending on the metric function $\Psi(r,\vartheta)$, where the underlying spin connection $\hat{\omega}_{1}$ associated with the KN geometry preserves its Minkowski values~(\ref{MinkowskiSP}). Thereby, the search of the non-propagating part of torsion with vanishing field strength tensors that satisfy the boundary conditions obtained in the spherically symmetric case is reduced to the obtainment of the rotated frame that fulfills the following equation:
\begin{equation}\label{vanishingantisymcurv}
\vartheta^{b}\,_{[\rho|}\mathcal{F}_{1}^{a}\,_{b|\mu\nu]}=0\,,
\end{equation}
with $\mathcal{F}_{1}^{a}\,_{b\mu\nu}=\partial_{\mu}\hat{\omega}_{1}^{a}\,_{b\nu}-\partial_{\nu}\hat{\omega}_{1}^{a}\,_{b\mu}+\hat{\omega}_{1}^{a}\,_{c\mu}\,\hat{\omega}_{1}^{c}\,_{b\nu}-\hat{\omega}_{1}^{a}\,_{c\nu}\,\hat{\omega}_{1}^{c}\,_{b\mu}$. The achievement of the complete solution requires then to find the additional propagating parts $\hat{\omega}_{2}$ and $\hat{\omega}_{3}$ of the spin connection, which provide nontrivial field strength tensors satisfying the field equations~(\ref{field_eq1}) and (\ref{field_eq2}).

\subsection{Rotating black hole solution}\label{subsec:axialB}

In general, a Lorentz matrix contains up to six degrees of freedom and can be expressed as a multiplication of six linear transformations, namely the Lorentz boosts and the space rotations:
\small{
\begin{align}\label{eq:lambda}
    \Lambda^a{}_b&=\left(
\begin{array}{cccc}
 1 & 0 & 0 & 0 \\
 0 & \cos\alpha_3 & -\sin\alpha_3 & 0 \\
 0 & \sin\alpha_3 & \cos\alpha_3 & 0 \\
 0 & 0 & 0 & 1 \\
\end{array}
\right)\times\left(
\begin{array}{cccc}
 1 & 0 & 0 & 0 \\
 0 & \cos\alpha_2 & 0 & -\sin\alpha_2 \\
 0 & 0 & 1 & 0 \\
 0 & \sin\alpha_2 & 0 & \cos\alpha_2 \\
\end{array}
\right)\times\left(
\begin{array}{cccc}
 1 & 0 & 0 & 0 \\
 0 & 1 & 0 & 0 \\
 0 & 0 & \cos\alpha_1 & -\sin\alpha_1 \\
 0 & 0 & \sin\alpha_1 & \cos\alpha_1 \\
\end{array}
\right)\nonumber\\
&\,\,\,\,\,\,\,\times \left(
\begin{array}{cccc}
 \cosh\beta_3 & 0 & 0 & \sinh\beta_3 \\
 0 & 1 & 0 & 0 \\
 0 & 0 & 1 & 0 \\
 \sinh\beta_3 & 0 & 0 & \cosh\beta_3 \\
\end{array}
\right)\times\left(
\begin{array}{cccc}
 \cosh\beta_2 & 0 & \sinh\beta_2 & 0 \\
 0 & 1 & 0 & 0 \\
 \sinh\beta_2 & 0 & \cosh\beta_2 & 0 \\
 0 & 0 & 0 & 1 \\
\end{array}
\right)\times\left(
\begin{array}{cccc}
 \cosh\beta_1 & \sinh\beta_1 & 0 & 0 \\
 \sinh\beta_1 & \cosh\beta_1 & 0 & 0 \\
 0 & 0 & 1 & 0 \\
 0 & 0 & 0 & 1 \\
\end{array}
\right)\,.
\end{align}}\normalsize
Accordingly, our construction requires the existence of certain values for the parameters $\{\alpha_{i}\}_{i=1}^{3}$ and $\{\beta_{i}\}_{i=1}^{3}$ which preserve the structure of the spin connection~(\ref{MinkowskiSP}) in a local frame for the non-propagating sector of the solution. In particular, such a Lorentz matrix can be split into a first transformation $\Lambda\big|_{\rm sph}$ parametrised by the Lorentz boost
\begin{equation}\label{tildebeta1}
    \tilde{\beta}_{1}=\arccosh\left(\frac{\Psi(r,\vartheta)+1}{2 \sqrt{\Psi(r,\vartheta)}}\right)\,,
\end{equation}
which fits the spherically symmetric parts of the vierbein and the spin connection with vanishing field strength tensors, and a second transformation $\Lambda\big|_{\rm axi}$ which fits the axisymmetric parts, namely
\begin{equation}
\Lambda^{a}\,_{b}=\Lambda^{a}\,_{c}\big|_{\rm axi}\Lambda^{c}\,_{b}\big|_{\rm sph}\,.
\end{equation}

Concerning the spherically symmetric branch, it is worthwhile to stress that any rotated basis provided by a Lorentz boost with an arbitrary value of $\beta_{1}$ is compatible with the spin connection~(\ref{MinkowskiSP}) and fulfills the constraint~(\ref{vanishingantisymcurv}). Nevertheless, by setting this parameter as (\ref{tildebeta1}) in the mentioned branch, the resulting tuple $\left(\vartheta^{a}\,_{\mu},\hat{\omega}_{1}^{a}\,_{b\mu}\right)$ respects the regularity of the torsion tensor \cite{Cembranos:2016gdt,Cembranos:2017pcs,Bahamonde:2020fnq}. On the other hand, the specific form of the axisymmetric part is subject to the resolution of the Expression~(\ref{vanishingantisymcurv}). The computation of this expression in terms of a general Lorentz matrix (\ref{eq:lambda}) gives rise to the following system of equations:
\begin{eqnarray}
    0&=&\cos \alpha_2 \sinh \beta_1 (\cos \alpha_1 \cosh \beta_2 \sinh \beta_3+\sin \alpha_1 \sinh \beta_2)+\sin \alpha_2 \cosh \beta_1\,,\\
    0&=&\sqrt{\left(r^2+a^2\cos ^2\vartheta\right)\Psi(r,\vartheta)+a^2 \sin ^2\vartheta} \,\bigl[\cos \alpha_2 \cosh \beta_1 \left(\cos \alpha_1 \cosh \beta_2 \sinh \beta_3+\sin \alpha_1 \sinh \beta_2\right)+\sin \alpha_2 \sinh \beta_1\bigr]\nonumber\\
    &&-\,a \sin \vartheta  \cos \alpha_1\cos \alpha_2 \cosh \beta_3\,,\\
    0&=&\sinh \beta_1 \bigl[\sinh \beta_2 \left(\cos \alpha_1 \cos \alpha_3-\sin \alpha_1 \sin \alpha_2 \sin \alpha_3\right)-\cosh \beta_2 \sinh \beta_3 \left(\cos \alpha_1 \sin \alpha_2 \sin \alpha_3+\sin \alpha_1 \cos \alpha_3\right)\bigr]\nonumber\\
    &&+\cos \alpha_2 \sin \alpha_3 \cosh \beta_1\,,\\
    0&=&\sqrt{\left(r^2+a^2 \cos ^2\vartheta\right)\Psi(r,\vartheta)+a^2 \sin ^2\vartheta } \, \big\{\cosh \beta_1 \bigl[\cosh \beta_2 \sinh \beta_3 \left(\cos \alpha_1 \sin \alpha_2 \sin \alpha_3+\sin \alpha_1 \cos \alpha_3\right)
    \bigr.
    \nonumber\\
    &&
    \bigl.
    +\sinh \beta_2 \left(\sin \alpha_1 \sin \alpha_2 \sin \alpha_3-\cos \alpha_1 \cos \alpha_3\right)\bigr]-\cos \alpha_2 \sin \alpha_3\sinh \beta_1\big\}\nonumber\\
    &&-\,a\sin \vartheta  \cosh \beta_3 (\cos \alpha_1 \sin \alpha_2 \sin \alpha_3+\sin \alpha_1 \cos \alpha_3)\,,
\end{eqnarray}
which turns out to be overdetermined. Nevertheless, if we set $\beta_{1}=\tilde{\beta}_{1}$ and $\alpha_1=\alpha_3=\beta_2=0$, then this system is reduced to the equations
\begin{eqnarray}
    0&=&\cos\alpha_2\sinh\beta_3\sinh\tilde{\beta}_{1}+\sin\alpha_2\cosh\tilde{\beta}_{1}\,,\\
    0&=&\sqrt{\left(r^2+a^2\cos^2\vartheta\right)\Psi(r,\vartheta)+a^2\sin ^2\vartheta}  \,\bigl(\cos\alpha_2\sinh\beta_3\cosh\tilde{\beta}_1+\sin\alpha_2\sinh\tilde{\beta}_1\bigr)-a \sin \vartheta  \cos \alpha_2 \cosh \beta_3\,,
\end{eqnarray}
which are solved by the parameters
\begin{align}
    \alpha_2&=\arccos\left[\frac{\sqrt{4(r^2+a^2\cos^2\vartheta)\Psi^2(r,\vartheta)-a^2(1-\Psi(r,\vartheta))^2\sin^2\vartheta}}{2\Psi(r,\vartheta)\sqrt{r^2+a^2\cos^2\vartheta}}\right]\,,\label{alpha2}\\
    \beta_3&=\arccosh\left[\frac{2\sqrt{(r^2+a^2\cos^2\vartheta)\Psi^{2}(r,\vartheta)+a^2\sin^2\vartheta\,\Psi(r,\vartheta)}}{\sqrt{4(r^2+a^2\cos^2\vartheta)\Psi^2(r,\vartheta)-a^2(1-\Psi(r,\vartheta))^2\sin^2\vartheta}}\right]\,.\label{beta3}
\end{align}
In any case, it is worthwhile to stress that the spin connection~(\ref{MinkowskiSP}) has a residual gauge invariance provided by the parameters $\alpha_{1}$ and $\beta_{1}$, in such a way that Eq.~(\ref{vanishingantisymcurv}) is preserved for any gauge transformation defined by them. Therefore, the Lorentz matrix that solves Eq.~\eqref{vanishingantisymcurv} can be described by the product of an axisymmetric part depending on the parameters (\ref{alpha2}) and (\ref{beta3}), and a spherically symmetric part given by the parameter (\ref{tildebeta1}):
\begin{align}\label{LorentzMatrixTransformation}
    \Lambda^a{}_b&=\left(
    \begin{array}{cccc}
    \cosh\beta_{3} & 0 & 0 & \sinh\beta_{3}\\
    -\sin\alpha_{2}\sinh\beta_{3} & \cos\alpha_{2} & 0 & -\sin\alpha_{2}\cosh\beta_{3}\\
    0 & 0 & 1 & 0\\
    \cos\alpha_{2}\sinh\beta_{3} & \sin\alpha_{2} & 0 & \cos\alpha_{2}\cosh\beta_{3}\\
    \end{array}
    \right)\times  \left(
    \begin{array}{cccc}
    \cosh\tilde{\beta}_1 & \sinh\tilde{\beta}_1 & 0 & 0\\
    \sinh\tilde{\beta}_1 & \cosh\tilde{\beta}_1 & 0 & 0\\
    0 & 0 & 1 & 0\\
    0 & 0 & 0 & 1\\
    \end{array}
    \right)\,.
\end{align}

The existence of this matrix allows us to set the rotated basis $\vartheta^{a}=\Lambda^{a}\,_{b}\,e^{b}$ where the first components of the spin connection of the solution acquire the Minkowski values~(\ref{MinkowskiSP}). In this sense, these components represent the Minkowski part of the spin connection, in such a way that all the field strength tensors of the model as well as the field equations (\ref{field_eq1}) and (\ref{field_eq2}) identically vanish if the spin connection is reduced to this part. The next step then consists of finding the propagating parts of the spin connection provided by the dynamical torsion and nonmetricity tensors. In the present case, the latter is reduced to a Weyl vector with both electric and magnetic components that preserve the invariance condition~(\ref{ICnonmetricity}), which sets such components into the trace parts of the spin connection:
\begin{equation}
    W_{\mu}=-\,\frac{1}{2}\,\hat{\omega}^{a}\,_{a\mu}\,,
\end{equation}
with $W_{\mu}=\left(w_{1}\left(r,\vartheta\right),w_{2}\left(r,\vartheta\right),w_{3}\left(r,\vartheta\right),w_{4}\left(r,\vartheta\right)\right)$. According to the decomposition (\ref{decomposition2}), any other contribution of this vector to the antisymmetric part of the spin connection in the rotated basis
\begin{equation}
    \hat{\omega}_{2}^{\left[a b\right]}\,_{\mu}=\vartheta^{a}\,_{\lambda}\,\vartheta^{b \rho}\mathring{K}_{2}^{\lambda}\,_{\rho\mu}-\frac{1}{4}\vartheta^{[a}\,_{\mu}\vartheta^{b]\nu}W_{\nu}\,,
\end{equation}
can be straightforwardly cancelled out by a non-propagating vector mode of torsion $\mathring{T}_{2}^{\nu}=(3/2)\,W^{\nu}$, in such a way that the dynamical degrees of freedom of nonmetricity are appropriately encoded in the mentioned trace. In agreement with the decomposition (\ref{curvaturedecomposition}), the antisymmetrized part of the curvature tensor then acquires a nonvanishing component $\mathring{R}^{\lambda}\,_{[\mu \nu \rho]}=\frac{1}{4}\tilde{R}^{\sigma}\,_{\sigma[\mu\nu}\delta_{\rho]}\,^{\lambda}$ and the dynamical contribution provided by the Weyl vector in the Lagrangian is reduced to an effective term $e_{1}/(16\pi)\tilde{R}^{\lambda}\,_{\lambda\mu\nu}\tilde{R}^{\rho}\,_{\rho}\,^{\mu\nu}$. This means that the symmetric component in the indices $\lambda$ and $\mu$ of Eq.~(\ref{field_eq2}) turns out to describe the propagation of the nonmetricity field in the present model and is reduced to the expression
\begin{equation}
    \nabla_{\mu}\tilde{R}^{\lambda}\,_{\lambda}\,^{\mu\nu}=0\,,\label{eqnon}
\end{equation}
which constitutes a Maxwell equation for the homothetic part of the curvature tensor. The procedure to find the corresponding solution can be applied as follows. First, we notice that despite of the cumbersome explicit form of the resulting equations in axial symmetry, they can be expressed in a relatively compact way that facilitates their resolution. Specifically, the components with $\nu=1$ and $\nu=2$ can be written as a combination involving the radial and polar components of the Weyl vector
\begin{eqnarray}
0&=&\left[\partial_{\vartheta}w_{2}(r,\vartheta)-\partial_{r}w_{3}(r,\vartheta)\right]F_1(r,\vartheta)+\partial_{\vartheta}\left[\partial_{\vartheta}w_{2}(r,\vartheta)-\partial_{r}w_{3}(r,\vartheta)\right]F_2(r,\vartheta)\,,\\
0&=&\left[\partial_{\vartheta}w_{2}(r,\vartheta)-\partial_{r}w_{3}(r,\vartheta)\right]F_3(r,\vartheta)+\partial_{r}\left[\partial_{\vartheta}w_{2}(r,\vartheta)-\partial_{r}w_{3}(r,\vartheta)\right]F_4(r,\vartheta)\,,
\end{eqnarray}
where $\{F_{i}(r,\vartheta)\}_{i=1}^{4}$ are functions including the parameter $a$ and $\Psi(r,\vartheta)$. Then, the following condition satisfies straightforwardly the mentioned equations:
\begin{equation}
    \partial_{\vartheta}w_{2}(r,\vartheta)=\partial_{r}w_{3}(r,\vartheta)\,.
\end{equation}
Note that this expression does not determine univocally the components of the Weyl vector, but one has to impose an additional constraint to set their values (e.g. a specific value for the quadratic scalar derived from this vector or for any of the mentioned components). For simplicity, here we consider the case where $w_{3}(r,\vartheta)=0$, which allows us to set the value of the radial component as a trivial generalisation of the spherically symmetric solution
\begin{equation}
    w_{2}(r,\vartheta)=-\,\frac{\kappa_{d,e}r}{(r^2+a^2\cos^{2}\vartheta)\Psi(r,\vartheta)+a^2\sin^{2}\vartheta}\,,
\end{equation}
where $\kappa_{d,e}$ is an integration constant. Additionally, the remaining equations with $\nu=0$ and $\nu=3$ can be expressed as

\begin{eqnarray}
0&=&G_{1}(r,\vartheta)\Psi(r,\vartheta)+G_{2}(r,\vartheta)\Psi^{2}(r,\vartheta)+G_{3}(r,\vartheta)+4au^{2}(r,\vartheta)\Psi^{3}(r,\vartheta)\left[a\,\partial_{rr}w_{1}(r,\vartheta)\sin^2\vartheta+\partial_{rr}w_{4}(r,\vartheta)\right]\nonumber\\
&&+\,4a\left[a\,\partial_{r}w_1(r,\vartheta)\sin^2\vartheta+\partial_{r}w_4(r,\vartheta)\right]\left[u(r,\vartheta)\Psi(r,\vartheta )+2a^2\sin ^2\vartheta\right]^{2}\partial_{r}\Psi(r,\vartheta)\sin ^2\vartheta\nonumber\\
&&+\,16\left(r^2+a^2\right)\partial_{\vartheta}\Psi(r,\vartheta)\left[\left(r^2+a^2\right)\partial_{\vartheta}w_{1}(r,\vartheta)+a\,\partial_{\vartheta}w_{4}(r,\vartheta)\right]\,,\label{non2}\\[1ex]
0&=&G_{4}(r,\vartheta)\Psi(r,\vartheta)+G_{5}(r,\vartheta)\Psi^{2}(r,\vartheta)+G_{6}(r,\vartheta)+64a\,\partial_{\vartheta}\Psi(r,\vartheta)\left[\left(r^2+a^2\right)\partial_{\vartheta}w_{1}(r,\vartheta)+a\,\partial_{\vartheta}w_{4}(r,\vartheta)\right]\nonumber\\
&&+\,16\left[a\,\partial_r w_1(r,\vartheta)\sin^2\vartheta+\partial_r w_4(r,\vartheta)\right]\left[u(r,\vartheta)\Psi(r,\vartheta)\csc\vartheta+2 a^2\sin\vartheta\right]^2\partial_{r}\Psi(r,\vartheta)\nonumber\\
&&+\,16u^2(r,\vartheta)\Psi^{3}(r,\vartheta)\left(a\,\partial_{rr}w_{1}(r,\vartheta)+\partial_{rr}w_{4}(r,\vartheta)\csc^{2}\vartheta\right)\,,\label{non3}
\end{eqnarray}
where $u(r,\vartheta)=2\left(r^2+a^2\cos^2\vartheta\right)$ and $\{G_{i}(r,\vartheta)\}_{i=1}^{6}$ are cumbersome expressions depending on $a$, $w_1(r,\vartheta)$ and $w_4(r,\vartheta)$. In order to solve these two equations, we demand from the time and azimuth components of the Weyl vector to be independent from the metric function $\Psi(r,\vartheta)$, which can be easily achieved by vanishing the terms related to $\partial_{\vartheta}\Psi(r,\vartheta),\, \partial_{r}\Psi(r,\vartheta)$ and $\Psi^3(r,\vartheta)$, namely
\begin{eqnarray}
    0&=&\left(r^2+a^2\right)\partial_{\vartheta}w_{1}(r,\vartheta)+a\,\partial_{\vartheta}w_{4}(r,\vartheta)\,,\label{ecweyl}\\
    0&=&a\,\partial_r w_1(r,\vartheta)\sin^2\vartheta+\partial_r w_4(r,\vartheta)\,.\label{eccweyl}
\end{eqnarray}
A straightforward integration of these expressions yields
\begin{eqnarray}
w_1(r,\vartheta)&=&w_{1b}(r)-\frac{aw_4(r,\vartheta)}{r^2+a^2}\,,\\
w_4(r,\vartheta)&=&\frac{\left(r^2+a^2\right) \left(w_{4b}(\vartheta)-2aw_{1b}(r)\sin^2\vartheta\right)}{2(r^2+a^2\cos^2\vartheta)}\,,
\end{eqnarray}
where $w_{1b}(r)$ and $w_{4b}(\vartheta)$ are integration functions to be determined by vanishing the rest of terms of the equations (\ref{non2}) and (\ref{non3}). By replacing these relations in the mentioned equations, we find that they are reduced to the following system of ordinary differential equations involving the integration functions:
\begin{eqnarray}
w_{1b}''(r)&=& \frac{aw_{4b}(\vartheta)+aw_{4b}'(\vartheta)\cot\vartheta-2rw_{1b}'(r)\left(r^2+a^2\cos(2\vartheta)\right)-a^{2}w_{1b}(r)(3+\cos(2\vartheta))}{\left(r^2+a^2\right)(r^2+a^2\cos^2\vartheta)}\,,\\
w_{4b}''(\vartheta)&=&\frac{4a\sin^2\vartheta\left[2(a^2-r^2)w_{1b}(r)-aw_{4b}(\vartheta )-2r\left(r^2+a^2\right)w_{1b}'(r)\right]+w_{4b}'(\vartheta)\left(2r^2+3a^2\cos(2\vartheta)-a^2\right)\cot\vartheta}{2 \left(r^2+a^2\cos^2\vartheta\right)}\,,\nonumber\\
\end{eqnarray}
which finally provides
\begin{eqnarray}
w_{1b}(r)=\frac{\kappa_{d,e}r-a\gamma\kappa_{d,m}}{r^2+a^2}\,,\quad w_{4b}(\vartheta)=2\kappa_{d,m}\left(\cos\vartheta-\gamma\right)\,,
\end{eqnarray}
where $\kappa_{d,m}$ and $\gamma$ are integration constants. Thereby, following this procedure the Maxwell equation (\ref{eqnon}) turns out to admit a type of monopole solution for the Weyl vector:
\begin{align}
    w_{1}(r,\vartheta)&=\frac{\kappa_{d,e}r-a\,\kappa_{d,m}\cos\vartheta}{r^2+a^2\cos^{2}\vartheta}\,, \quad w_{2}(r,\vartheta)=-\,\frac{\kappa_{d,e}r}{(r^2+a^2\cos^{2}\vartheta)\Psi(r,\vartheta)+a^2\sin^{2}\vartheta}\,,\nonumber\\
    w_{3}(r,\vartheta)&=0\,, \quad w_{4}(r,\vartheta)=\kappa_{d,m}\left(\frac{r^2+a^2}{r^2+a^2\cos^{2}\vartheta}\cos\vartheta-\gamma\right)-a\,\frac{\kappa_{d,e}r\sin^{2}\vartheta}{r^2+a^2\cos^{2}\vartheta}\,.
\end{align}
The integration constants $\kappa_{d,e}$ and $\kappa_{d,m}$ represent then electric and magnetic dilation charges, respectively. The additional constant $\gamma$ can be normalised as $\gamma=\pm 1$ for the corresponding northern and southern hemispheres surrounding the monopole, given the fact that the contribution of the Weyl vector in the trace part of the spin connection is defined up to a gauge transformation and the regularity of magnetic monopoles under a topological description \cite{wu1976dirac,Shnir_2005}. On the other hand, it is straightforward to note that the antisymmetric part of the field equation~(\ref{field_eq2}) is not in general reduced to a simple Maxwell equation for the torsion sector but contains additional nonlinear terms that in the axisymmetric case require a nontrivial coupling between the spin charge $\kappa_{s}$ and the orbital angular momentum $a$ in the solution to fulfill the equation. In the simplest scenario defined as the decoupling limit of these two quantities, they are constrained by the relation $|a\kappa_{s}| \ll 1$ and a Coulomb-like dynamical part of the torsion tensor
\begin{align}
    \bar{T}^{\vartheta}\,_{\varphi t}=-\,\bar{T}^{\varphi}\,_{\vartheta t}\sin^{2}\vartheta=-\,\bar{T}^{\vartheta}\,_{\varphi r}\Psi(r,\vartheta)=\bar{T}^{\varphi}\,_{\vartheta r}\Psi(r,\vartheta)\sin^{2}\vartheta=\frac{\kappa_{s}\sin\vartheta}{r}+\mathcal{O}(a\kappa_{s})\,,
\end{align}
provides a Maxwell equation for the corresponding field strength tensors
\begin{eqnarray}
\nabla_{\lambda}\tilde{R}^{\lambda}\,_{[\rho \mu \nu]}&=&0\label{Maswell1}\,,\\
\nabla_{\mu}\tilde{R}^{[\mu\nu]}&=&0\label{Maswell2}\,,
\end{eqnarray}
together with the closure conditions
\begin{eqnarray}
\nabla_{[\sigma}\tilde{\mathcal{R}}_{\lambda\rho\mu\nu]}&=&0\label{closure1}\,,\\
\nabla_{[\lambda}\tilde{\mathcal{R}}_{\mu\nu]}&=&0\label{closure2}\,,
\end{eqnarray}
with $\tilde{\mathcal{R}}_{\lambda\rho\mu\nu} \equiv \tilde{R}_{\lambda[\rho \mu \nu]}$ and $\tilde{\mathcal{R}}_{\mu\nu} \equiv \tilde{R}_{[\mu\nu]}$. Then, the resulting terms of Eq.~(\ref{field_eq2}) can be expressed as powers of $a\kappa_{s}$ and consistently dismissed if $|a\kappa_{s}| \ll 1$. The remaining field equation~(\ref{field_eq1}) sets the metric function $\Psi(r,\vartheta)$ relating the constant parameter $k$ and the hypermomentum charges as follows:
\begin{equation}
    \Psi(r,\vartheta)=1-\frac{2mr-\bigl[d_{1}\kappa^{2}_{s}-4e_{1}(\kappa^{2}_{d,e}+\kappa^{2}_{d,m})\bigr]}{r^{2}+a^{2}\cos^{2}\vartheta}\,.
\end{equation}

Collecting all the parts of the solution, the full anholonomic connection expressed in the local frame provided by the Lorentz transformation (\ref{LorentzMatrixTransformation}) turns out to acquire the following form:
\begin{equation}
    \hat{\omega}=\sum_{i=1}^{3}\hat{\omega}_{i}\,,
\end{equation}
with
\begin{eqnarray}
    \hat{\omega}_{1}&=&\frac{1}{2}\bigl(L_{[\hat{1}\hat{2}]}-L_{[\hat{0}\hat{2}]}\bigr)\,d\vartheta +\frac{1}{2}\sin\vartheta \left[\bigl(L_{[\hat{1}\hat{3}]}-L_{[\hat{0}\hat{3}]}\bigr)+2\cot\vartheta L_{[\hat{2}\hat{3}]}\right] d\varphi\,,\\
    \hat{\omega}_{2}&=&\frac{1}{2}\,\eta^{ab}L_{ab}
\biggl\{
-\,\frac{\kappa_{d,e}r-a\kappa_{d,m}\cos\vartheta}{r^2+a^2\cos^{2}\vartheta}\,dt+\frac{a\kappa_{d,e}r\sin^{2}\vartheta+\kappa_{d,m}\left[\pm\left(r^2+a^{2}\cos^{2}\vartheta\right)-\left(r^2+a^2\right)\cos\vartheta\right]}{r^2+a^2\cos^{2}\vartheta}\,d\varphi
\biggr.
\nonumber\\
\biggl.
&&+\,\frac{\kappa_{d,e}r}{(r^2+a^2\cos^{2}\vartheta)\Psi(r,\vartheta)+a^2\sin^{2}\vartheta}\,dr
\biggr\}\,,\\
    \hat{\omega}_{3}&=&-\,\frac{\kappa_{s}}{r}\left[dt-\frac{dr}{\Psi(r,\vartheta)}\right]L_{[\hat{2}\hat{3}]}\,,
\end{eqnarray}
where $L_{a b}$ are the generators of the group $GL(4,R)$ and $\eta^{ab}=\textrm{diag}\left(1,-\,1,-\,1,-\,1\right)$. The non-propagating irreducible modes of torsion can then be expressed as the sum of a first set of pieces related to vanishing field strength tensors
\begin{eqnarray}
    \mathring{T}_{1\mu}&=&2\vartheta_{a}\,^{\lambda}\left(\partial_{[\lambda}\vartheta^{a}\,_{\mu]}+\hat{\omega}^{a}_{1}\,_{b[\lambda}\,\vartheta^{b}\,_{\mu]}\right)\,,\\
    \mathring{S}_{1\mu}&=&2\varepsilon_{\mu\lambda}\,^{\nu\rho}\vartheta_{a}\,^{\lambda}\left(\partial_{[\rho}\vartheta^{a}\,_{\nu]}+\hat{\omega}^{a}_{1}\,_{b[\rho}\,\vartheta^{b}\,_{\nu]}\right)\,,\\
    \mathring{t}_{1\lambda\mu\nu}&=&\frac{2}{3}\,\eta_{ab}\Bigl\{2\vartheta^{a}\,_{\lambda}\left(\partial_{[\nu}\vartheta^{b}\,_{\mu]}+\hat{\omega}^{b}_{1}\,_{c[\nu}\,\vartheta^{c}\,_{\mu]}\right)+\vartheta^{a}\,_{\mu}\left(\partial_{[\nu}\vartheta^{b}\,_{\lambda]}+\hat{\omega}^{b}_{1}\,_{c[\nu}\,\vartheta^{c}\,_{\lambda]}\right)-\,\vartheta^{a}\,_{\nu}\left(\partial_{[\mu}\vartheta^{b}\,_{\lambda]}+\hat{\omega}^{b}_{1}\,_{c[\mu}\,\vartheta^{c}\,_{\lambda]}\right)
    \Bigr.
    \nonumber\\
    \Bigl.
    &&+\,g^{\sigma\rho}\vartheta^{a}\,_{\sigma}\bigl[g_{\lambda\mu}\left(\partial_{[\rho}\vartheta^{b}\,_{\nu]}+\hat{\omega}^{b}_{1}\,_{c[\rho}\,\vartheta^{c}\,_{\nu]}\right)-g_{\lambda\nu}\left(\partial_{[\rho}\vartheta^{b}\,_{\mu]}+\hat{\omega}^{b}_{1}\,_{c[\rho}\,\vartheta^{c}\,_{\mu]}\right)\bigr]\Bigr\}\,,
\end{eqnarray}
and an additional vector mode
\begin{equation}
    \mathring{T}_{2\mu}=-\,\frac{3}{4}\,\hat{\omega}^{a}_{2}\,_{a\mu}\,,
\end{equation}
which gives rise to a degenerate homothetic component in the curvature tensor. On the other hand, the propagating modes of the torsion tensor can be written as
\begin{eqnarray}
    \bar{S}_{\mu}&=&2\varepsilon_{\mu\lambda}\,^{\nu\rho}\vartheta_{a}\,^{\lambda}\left(\partial_{[\rho}\vartheta^{a}\,_{\nu]}+\hat{\omega}^{a}_{3}\,_{b[\rho}\,\vartheta^{b}\,_{\nu]}\right)\,,\\
    \bar{t}_{\lambda\mu\nu}&=&\frac{2}{3}\,\eta_{ab}\left[2\vartheta^{a}\,_{\lambda}\left(\partial_{[\nu}\vartheta^{b}\,_{\mu]}+\hat{\omega}^{b}_{3}\,_{c[\nu}\,\vartheta^{c}\,_{\mu]}\right)+\vartheta^{a}\,_{\mu}\left(\partial_{[\nu}\vartheta^{b}\,_{\lambda]}+\hat{\omega}^{b}_{3}\,_{c[\nu}\,\vartheta^{c}\,_{\lambda]}\right)-\,\vartheta^{a}\,_{\nu}\left(\partial_{[\mu}\vartheta^{b}\,_{\lambda]}+\hat{\omega}^{b}_{3}\,_{c[\mu}\,\vartheta^{c}\,_{\lambda]}\right)\right]\,.
\end{eqnarray}
All these pieces provide the following field strength tensors of the post-Riemannian gravitational interaction:
\begin{eqnarray}
\tilde{R}^{\sigma}\,_{\sigma\mu\nu}&=&4\nabla_{[\nu}W_{\mu]}\,,\\
\tilde{R}^{\lambda}\,_{[\mu\nu\rho]}&=&\frac{1}{4}\tilde{R}^{\sigma}\,_{\sigma[\mu\nu}\delta_{\rho]}\,^{\lambda}+\bar{R}^{\lambda}\,_{[\mu\nu\rho]}\,,\\
\tilde{R}_{[\mu\nu]}&=&\frac{1}{4}\tilde{R}^{\sigma}\,_{\sigma\mu\nu}+\bar{R}_{[\mu\nu]}\,,\\
\bar{R}^{\lambda}\,_{[\mu \nu \rho]}&=&\frac{1}{6}\varepsilon^{\lambda}\,_{\sigma[\rho\nu}\nabla_{\mu]}\bar{S}^{\sigma}+\nabla_{[\mu}\bar{t}^{\lambda}\,_{\rho\nu]}+\frac{1}{4}\varepsilon^{\lambda}\,_{\omega\sigma[\rho}\mathring{t}_{1}^{\sigma}\,_{\mu\nu]}\bar{S}^{\omega}-\frac{1}{18}\varepsilon_{\sigma\mu\nu\rho}\mathring{T}_{1}^{\lambda}\bar{S}^{\sigma}\,,\\
\bar{R}_{[\mu\nu]}&=&\frac{1}{12}\varepsilon^{\lambda}\,_{\sigma\mu\nu}\nabla_{\lambda}\bar{S}^{\sigma}+\frac{1}{2}\nabla_{\lambda}\bar{t}^{\lambda}\,_{\mu\nu}\,.
\end{eqnarray}

Any nonlinearity arising from the coupling between the orbital and the spin angular momentum is expected to generate deviations from the expressions (\ref{Maswell1})-(\ref{closure2}) and require further corrections in the geometry of the space-time. In any case, it is worthwhile to stress that the aforementioned expressions do not involve the fulfillment of the field equations of the model, but they constitute a particular property which is realised by the solution in the decoupling limit.

Additionally, the solution can be generalised to include a nonvanishing cosmological constant $\Lambda$ and electromagnetic fields with electric and magnetic charges $q_{e}$ and $q_{m}$, which at the same time are decoupled from the general linear connection under the assumption of the minimal coupling principle. Such a natural extension can be obtained in the limit $|a\kappa_{s}| \ll 1$ by modifying the Lorentz parameters, as well as the metric tensor and the Weyl vector as follows:
\begin{align}
    \alpha_2&=\arccos\left[\frac{\sqrt{4(r^2+a^2\cos^2\vartheta)\Psi^2(r,\vartheta)-a^2(1-\Psi(r,\vartheta))^2\sin^2\vartheta-\frac{1}{3}a^4\Lambda\left(1+\Psi(r,\vartheta)\right)^2\sin^2\vartheta\cos^2\vartheta}}{2\sqrt{\left(r^2+a^2\cos^2\vartheta\right)\Psi^{2}(r,\vartheta)-\frac{1}{3}a^4\Lambda\Psi(r,\vartheta)\sin^2\vartheta\cos^2\vartheta}}\right]\,,\\
  \beta_3&=\arccosh\left[\frac{2\sqrt{(r^2+a^2\cos^2\vartheta)\Psi^{2}(r,\vartheta)+a^2\sin^2\vartheta\,\Psi(r,\vartheta)}}{\sqrt{4(r^2+a^2\cos^2\vartheta)\Psi^2(r,\vartheta)-a^2(1-\Psi(r,\vartheta))^2\sin^2\vartheta-\frac{1}{3}a^4\Lambda\left(1+\Psi(r,\vartheta)\right)^{2}\sin^2\vartheta\cos^2\vartheta}}\right]\,,
\end{align}
\begin{align}
    w_{1}(r,\vartheta)&=\frac{\kappa_{d,e}r-a\,\kappa_{d,m}\cos\vartheta}{r^2+a^2\cos^{2}\vartheta}\,, \quad w_{2}(r,\vartheta)=-\,\frac{\kappa_{d,e}r}{(r^2+a^2\cos^{2}\vartheta)\Psi(r,\vartheta)+a^2\sin^{2}\vartheta}\,,\nonumber\\
    w_{3}(r,\vartheta)&=0\,, \quad w_{4}(r,\vartheta)=\kappa_{d,m}\left(\frac{r^2+a^2}{r^2+a^2\cos^{2}\vartheta}\cos\vartheta-\gamma\right)-a\,\frac{\kappa_{d,e}r\sin^{2}\vartheta}{r^2+a^2\cos^{2}\vartheta}\,,
\end{align}
\begin{eqnarray}
ds^2&=&\Big(\Psi(r,\vartheta)-\frac{1}{3}\frac{a^4\Lambda\cos^2\vartheta\sin^2\vartheta}{r^2+a^2\cos^2\vartheta}\Big)\,dt^2
-\frac{r^2+a^2\cos^{2}\vartheta}{\left(r^2+a^2\cos^{2}\vartheta\right)\Psi(r,\vartheta)+a^2\sin^{2}\vartheta}\,dr^2-\frac{r^2+a^2\cos^2\vartheta}{1+\frac{1}{3}a^2\Lambda\cos^2\vartheta}\,d\vartheta^2\nonumber\\
&&-\,\sin^2\vartheta\left[r^2+a^2+a^2(1-\Psi(r,\vartheta))\sin ^2\vartheta+\frac{a^2 \Lambda\cos^2\vartheta\left(r^2+a^2\right)^2}{3 \left(r^2+a^2\cos^2\vartheta\right)}\right]d\varphi^2\nonumber\\
&&+\,2a\sin^2\vartheta\left[1-\Psi(r,\vartheta )+\frac{a^2\Lambda\left(r^2+a^2\right)\cos^2\vartheta}{3\left(r^2+a^2\cos^2\vartheta\right)}\right]dtd\varphi\,,
\end{eqnarray} with $\Psi(r,\vartheta)=1-\big\{2mr-\bigl[q_{e}^{2}+q_{m}^{2}+d_{1}\kappa^{2}_{s}-4e_{1}(\kappa^{2}_{d,e}+\kappa^{2}_{d,m})\bigr]+\frac{1}{3}\Lambda r^2(r^2+a^2)\big\}/(r^{2}+a^{2}\cos^{2}\vartheta)$. As can be seen, the corrections provided by the torsion and nonmetricity tensors act in the metric tensor as the ones depending on the electromagnetic charges and it is possible to collect all the contributions along with the cosmological constant within a common space-time. Nevertheless, the balance between the hypermomentum charges is not restricted to any particular constraint and remarkably a branch with a negative effective correction of the KN geometry may occur and cancel out the inner Cauchy horizon in the presence of torsion and nonmetricity, in contrast to the ordinary counterpart of GR \cite{Bahamonde:2021akc}.

\section{Conclusions}\label{sec:conclusions}

In the present work, we have analysed axial symmetry in the framework of MAG and obtained a new rotating black hole solution with dynamical torsion and nonmetricity fields. Indeed, we have shown that despite the highly nonlinear character of the field equations and the number of unknown axisymmetric components of the metric, torsion and nonmetricity tensors, it is possible to perform a decomposition into propagating and non-propagating modes to strongly simplify the problem and reach the solution. This procedure contrasts with different methods successfully applied in previous literature to generate rotating black hole solutions in GR and other theories of gravity from their corresponding nonrotating counterparts, such as the Newman-Janis algorithm \cite{Newman:1965tw,Erbin:2014aya}, which in the present framework do not provide a solution since both the tetrad field and the anholonomic connection of the group $GL(4,R)$ must be extended in an independent way.

The resulting configuration describes a KN metric-affine geometry sourced by electric and magnetic dilation charges related to the nonmetricity tensor, as well as by a spin charge related to the torsion tensor. The latter is constrained by the relation $|a\kappa_{s}| \ll 1$, which characterises the decoupling limit between the orbital and the spin angular momentum of the solution. In this sense, from a phenomenological point of view, the solution can be used to represent the asymptotic gravitational field of a rotating black hole or compact star formed in a metric-affine regime with scale invariance \cite{Bahamonde:2021akc}, whose spin charge is much lower than the external rotation, or vice versa (i.e. in such a way that the constraint $|a\kappa_{s}| \ll 1$ holds). As is shown, the dynamics of torsion and nonmetricity alters the geometry of the space-time, which means that additional modifications provided by a strong coupling between the orbital and the spin angular momentum are expected to appear for unconstrained values of $a$ and $\kappa_s$, even in the Riemannian sector described by the metric tensor. In fact, the existing correspondence between the geometry of the space-time and the hypermomentum of matter sets the framework of MAG as the main candidate to describe a gravitational spin-orbit interaction beyond the standard approach of GR. Research on the strong coupling regime beyond the KN space-time is currently underway.

On the other hand, we also obtain the generalised Kerr-Newman-de Sitter solution in the presence of a cosmological constant and external electromagnetic fields, by analogy with the standard case. In this sense, a canonical metric structure for the most general type D solution of the Einstein-Maxwell model depending on the NUT charge and the acceleration parameter is also expected to hold in the present case \cite{Griffiths:2009dfa,Podolsky:2021zwr}. Further research following these lines will be addressed in future works.

\bigskip
\bigskip
\noindent
\section*{Acknowledgements}
S.B. was supported by the Estonian Research Council grants PRG356 ``Gauge Gravity"  and by the European Regional Development Fund through the Center of Excellence TK133 ``The Dark Side of the Universe". J.G.V. was supported by the European Regional Development Fund and the programme Mobilitas Pluss (Grant No. MOBJD541).
\newpage

\bibliographystyle{utphys}
\bibliography{references}

\providecommand{\href}[2]{#2}\begingroup\raggedright\begin{thebibliography}{10}

\bibitem{Akiyama:2019cqa}
{K. Akiyama et al. (The Event Horizon Telescope Collaboration)}, ``{First M87
  Event Horizon Telescope Results. I. The Shadow of the Supermassive Black
  Hole},'' \href{http://dx.doi.org/10.3847/2041-8213/ab0ec7}{{\em Astrophys.
  J.} {\bf 875} (2019) no.~1, L1},
\href{http://arxiv.org/abs/1906.11238}{{\tt arXiv:1906.11238 [astro-ph.GA]}}.

\bibitem{Abbott:2016blz}
{B. P. Abbott et al. (LIGO Scientific and Virgo Collaborations)},
  ``{Observation of Gravitational Waves from a Binary Black Hole Merger},''
  \href{http://dx.doi.org/10.1103/PhysRevLett.116.061102}{{\em Phys. Rev.
  Lett.} {\bf 116} (2016) no.~6, 061102},
\href{http://arxiv.org/abs/1602.03837}{{\tt arXiv:1602.03837 [gr-qc]}}.

\bibitem{TheLIGOScientific:2017qsa}
{B. P. Abbott et al. (LIGO Scientific and Virgo Collaborations)}, ``{GW170817:
  Observation of Gravitational Waves from a Binary Neutron Star Inspiral},''
  \href{http://dx.doi.org/10.1103/PhysRevLett.119.161101}{{\em Phys. Rev.
  Lett.} {\bf 119} (2017) no.~16, 161101},
  \href{http://arxiv.org/abs/1710.05832}{{\tt arXiv:1710.05832 [gr-qc]}}.

\bibitem{LIGOScientific:2021qlt}
{B. P. Abbott et al. (LIGO Scientific, KAGRA and Virgo Collaborations)},
  ``{Observation of Gravitational Waves from Two Neutron Star\textendash{}Black
  Hole Coalescences},'' \href{http://dx.doi.org/10.3847/2041-8213/ac082e}{{\em
  Astrophys. J. Lett.} {\bf 915} (2021) no.~1, L5},
  \href{http://arxiv.org/abs/2106.15163}{{\tt arXiv:2106.15163 [astro-ph.HE]}}.

\bibitem{Kerr:1963ud}
R.~P. Kerr, ``{Gravitational field of a spinning mass as an example of
  algebraically special metrics},''
  \href{http://dx.doi.org/10.1103/PhysRevLett.11.237}{{\em Phys. Rev. Lett.}
  {\bf 11} (1963)  237--238}.

\bibitem{Broderick:2013rlq}
A.~E. Broderick, T.~Johannsen, A.~Loeb, and D.~Psaltis, ``{Testing the No-Hair
  Theorem with Event Horizon Telescope Observations of Sagittarius A*},''
  \href{http://dx.doi.org/10.1088/0004-637X/784/1/7}{{\em Astrophys. J.} {\bf
  784} (2014)  7}, \href{http://arxiv.org/abs/1311.5564}{{\tt arXiv:1311.5564
  [astro-ph.HE]}}.

\bibitem{Cardoso:2016ryw}
V.~Cardoso and L.~Gualtieri, ``{Testing the black hole
  \textquoteleft{}no-hair\textquoteright{} hypothesis},''
  \href{http://dx.doi.org/10.1088/0264-9381/33/17/174001}{{\em Class. Quant.
  Grav.} {\bf 33} (2016) no.~17, 174001},
  \href{http://arxiv.org/abs/1607.03133}{{\tt arXiv:1607.03133 [gr-qc]}}.

\bibitem{Isi:2019aib}
M.~Isi, M.~Giesler, W.~M. Farr, M.~A. Scheel, and S.~A. Teukolsky, ``{Testing
  the no-hair theorem with GW150914},''
  \href{http://dx.doi.org/10.1103/PhysRevLett.123.111102}{{\em Phys. Rev.
  Lett.} {\bf 123} (2019) no.~11, 111102},
  \href{http://arxiv.org/abs/1905.00869}{{\tt arXiv:1905.00869 [gr-qc]}}.

\bibitem{Johannsen:2015hib}
T.~Johannsen, A.~E. Broderick, P.~M. Plewa, S.~Chatzopoulos, S.~S. Doeleman,
  F.~Eisenhauer, V.~L. Fish, R.~Genzel, O.~Gerhard, and M.~D. Johnson,
  ``{Testing General Relativity with the Shadow Size of Sgr A*},''
  \href{http://dx.doi.org/10.1103/PhysRevLett.116.031101}{{\em Phys. Rev.
  Lett.} {\bf 116} (2016) no.~3, 031101},
  \href{http://arxiv.org/abs/1512.02640}{{\tt arXiv:1512.02640 [astro-ph.GA]}}.

\bibitem{Visser:2007fj}
M.~Visser, ``{The Kerr spacetime: A brief introduction},'' in {\em {Kerr Fest:
  Black Holes in Astrophysics, General Relativity and Quantum Gravity}}.
\newblock 6, 2007.
\newblock \href{http://arxiv.org/abs/0706.0622}{{\tt arXiv:0706.0622 [gr-qc]}}.

\bibitem{Bahamonde:2020snl}
S.~Bahamonde, J.~Gigante~Valcarcel, L.~J\"arv, and C.~Pfeifer, ``{Exploring
  Axial Symmetry in Modified Teleparallel Gravity},''
  \href{http://dx.doi.org/10.1103/PhysRevD.103.044058}{{\em Phys. Rev. D} {\bf
  103} (2021) no.~4, 044058}, \href{http://arxiv.org/abs/2012.09193}{{\tt
  arXiv:2012.09193 [gr-qc]}}.

\bibitem{Planck:2018vyg}
{N. Aghanim et al. (Planck Collaboration)}, ``{Planck 2018 results. VI.
  Cosmological parameters},''
  \href{http://dx.doi.org/10.1051/0004-6361/201833910}{{\em Astron. Astrophys.}
  {\bf 641} (2020)  A6}, \href{http://arxiv.org/abs/1807.06209}{{\tt
  arXiv:1807.06209 [astro-ph.CO]}}. [Erratum: Astron.Astrophys. 652, C4
  (2021)].

\bibitem{Wong:2019kwg}
{K. C. Wong et al.}, ``{H0LiCOW \textendash{} XIII. A 2.4 per cent measurement
  of \ensuremath{H_0} from lensed quasars: 5.3\ensuremath{\sigma} tension
  between early- and late-Universe probes},''
  \href{http://dx.doi.org/10.1093/mnras/stz3094}{{\em Mon. Not. Roy. Astron.
  Soc.} {\bf 498} (2020) no.~1, 1420--1439},
  \href{http://arxiv.org/abs/1907.04869}{{\tt arXiv:1907.04869 [astro-ph.CO]}}.

\bibitem{DiValentino:2021izs}
E.~Di~Valentino, O.~Mena, S.~Pan, L.~Visinelli, W.~Yang, A.~Melchiorri, D.~F.
  Mota, A.~G. Riess, and J.~Silk, ``{In the realm of the Hubble
  tension\textemdash{}a review of solutions},''
  \href{http://dx.doi.org/10.1088/1361-6382/ac086d}{{\em Class. Quant. Grav.}
  {\bf 38} (2021) no.~15, 153001}, \href{http://arxiv.org/abs/2103.01183}{{\tt
  arXiv:2103.01183 [astro-ph.CO]}}.

\bibitem{Peebles:2002gy}
P.~J.~E. Peebles and B.~Ratra, ``{The Cosmological Constant and Dark Energy},''
  \href{http://dx.doi.org/10.1103/RevModPhys.75.559}{{\em Rev. Mod. Phys.} {\bf
  75} (2003)  559--606}, \href{http://arxiv.org/abs/astro-ph/0207347}{{\tt
  arXiv:astro-ph/0207347}}.

\bibitem{Navarro:1995iw}
J.~F. Navarro, C.~S. Frenk, and S.~D.~M. White, ``{The Structure of Cold Dark
  Matter Halos},'' \href{http://dx.doi.org/10.1086/177173}{{\em Astrophys. J.}
  {\bf 462} (1996)  563--575},
  \href{http://arxiv.org/abs/astro-ph/9508025}{{\tt arXiv:astro-ph/9508025}}.

\bibitem{Nojiri:2017ncd}
S.~Nojiri, S.~D. Odintsov, and V.~K. Oikonomou, ``{Modified Gravity Theories on
  a Nutshell: Inflation, Bounce and Late-time Evolution},''
  \href{http://dx.doi.org/10.1016/j.physrep.2017.06.001}{{\em Phys. Rept.} {\bf
  692} (2017)  1--104}, \href{http://arxiv.org/abs/1705.11098}{{\tt
  arXiv:1705.11098 [gr-qc]}}.

\bibitem{Joyce:2014kja}
A.~Joyce, B.~Jain, J.~Khoury, and M.~Trodden, ``{Beyond the Cosmological
  Standard Model},''
  \href{http://dx.doi.org/10.1016/j.physrep.2014.12.002}{{\em Phys. Rept.} {\bf
  568} (2015)  1--98}, \href{http://arxiv.org/abs/1407.0059}{{\tt
  arXiv:1407.0059 [astro-ph.CO]}}.

\bibitem{CANTATA:2021ktz}
{E. N. Saridakis et al. (CANTATA Collaboration)}, ``{Modified Gravity and
  Cosmology: An Update by the CANTATA Network},''
  \href{http://arxiv.org/abs/2105.12582}{{\tt arXiv:2105.12582 [gr-qc]}}.

\bibitem{Hehl:1994ue}
F.~W. Hehl, J.~D. McCrea, E.~W. Mielke, and Y.~Ne'eman, ``{Metric-Affine gauge
  theory of gravity: Field equations, Noether identities, world spinors, and
  breaking of dilation invariance},''
  \href{http://dx.doi.org/10.1016/0370-1573(94)00111-F}{{\em Phys. Rept.} {\bf
  258} (1995)  1--171},
\href{http://arxiv.org/abs/gr-qc/9402012}{{\tt arXiv:gr-qc/9402012 [gr-qc]}}.

\bibitem{Blagojevic:2013xpa}
M.~Blagojevi\'c and F.~W. Hehl, eds., {\em {Gauge Theories of Gravitation}: {A
  Reader with Commentaries}}.
\newblock World Scientific, Singapore, 2013.

\bibitem{Cabral:2020fax}
F.~Cabral, F.~S. Lobo, and D.~Rubiera-Garcia, ``{Fundamental Symmetries and
  Spacetime Geometries in Gauge Theories of Gravity: Prospects for Unified
  Field Theories},'' \href{http://dx.doi.org/10.3390/universe6120238}{{\em
  Universe} {\bf 6} (2020) no.~12, 238},
  \href{http://arxiv.org/abs/2012.06356}{{\tt arXiv:2012.06356 [gr-qc]}}.

\bibitem{Vlachynsky:1996zh}
E.~J. Vlachynsky, R.~Tresguerres, Y.~N. Obukhov, and F.~W. Hehl, ``{An axially
  symmetric solution of metric-affine gravity},''
  \href{http://dx.doi.org/10.1088/0264-9381/13/12/016}{{\em Class. Quant.
  Grav.} {\bf 13} (1996)  3253--3260},
  \href{http://arxiv.org/abs/gr-qc/9604035}{{\tt arXiv:gr-qc/9604035}}.

\bibitem{Hehl:1999sb}
F.~W. Hehl and A.~Macias, ``{Metric-Affine gauge theory of gravity. 2. Exact
  solutions},'' \href{http://dx.doi.org/10.1142/S0218271899000316}{{\em Int. J.
  Mod. Phys. D} {\bf 8} (1999)  399--416},
  \href{http://arxiv.org/abs/gr-qc/9902076}{{\tt arXiv:gr-qc/9902076}}.

\bibitem{Baekler:2006de}
P.~Baekler and F.~W. Hehl, ``{Rotating black holes in metric-affine gravity},''
  \href{http://dx.doi.org/10.1142/S0218271806008589}{{\em Int. J. Mod. Phys. D}
  {\bf 15} (2006)  635--668}, \href{http://arxiv.org/abs/gr-qc/0601063}{{\tt
  arXiv:gr-qc/0601063}}.

\bibitem{Bahamonde:2020fnq}
S.~Bahamonde and J.~G. Valcarcel, ``{New models with independent dynamical
  torsion and nonmetricity fields},''
  \href{http://dx.doi.org/10.1088/1475-7516/2020/09/057}{{\em JCAP} {\bf 09}
  (2020)  057}, \href{http://arxiv.org/abs/2006.06749}{{\tt arXiv:2006.06749
  [gr-qc]}}.

\bibitem{Cembranos:2016gdt}
J.~A.~R. Cembranos and J.~G. Valcarcel, ``{New torsion black hole solutions in
  Poincar\'e gauge theory},''
  \href{http://dx.doi.org/10.1088/1475-7516/2017/01/014}{{\em JCAP} {\bf 01}
  (2017)  014}, \href{http://arxiv.org/abs/1608.00062}{{\tt arXiv:1608.00062
  [gr-qc]}}.

\bibitem{Cembranos:2017pcs}
J.~A.~R. Cembranos and J.~Gigante~Valcarcel, ``{Extended Reissner--Nordström
  solutions sourced by dynamical torsion},''
  \href{http://dx.doi.org/10.1016/j.physletb.2018.01.081}{{\em Phys. Lett. B}
  {\bf 779} (2018)  143--150}, \href{http://arxiv.org/abs/1708.00374}{{\tt
  arXiv:1708.00374 [gr-qc]}}.

\bibitem{Blagojevic:2021pqp}
M.~Blagojevi\'c and B.~Cvetkovi\'c, ``{Entropy of Reissner-Nordstr\"om-like
  black holes},'' \href{http://dx.doi.org/10.1016/j.physletb.2021.136815}{{\em
  Phys. Lett. B} {\bf 824} (2022)  136815},
  \href{http://arxiv.org/abs/2112.02099}{{\tt arXiv:2112.02099 [gr-qc]}}.

\bibitem{Obukhov:2020hlp}
Y.~N. Obukhov, ``{Generalized Birkhoff theorem in the Poincar\'e gauge gravity
  theory},'' \href{http://dx.doi.org/10.1103/PhysRevD.102.104059}{{\em Phys.
  Rev. D} {\bf 102} (2020) no.~10, 104059},
  \href{http://arxiv.org/abs/2009.00284}{{\tt arXiv:2009.00284 [gr-qc]}}.

\bibitem{Neeman:1977iup}
Y.~Ne'eman, ``{Spinor Type Fields with Linear, Affine and General Coordinate
  Transformations},'' {\em Ann. Inst. H. Poincar\'{e} Phys. Theor.} {\bf 28}
  (1978)  369--378.

\bibitem{Neeman:1987pzd}
Y.~Ne'eman and D.~Sijacki, ``{$\overline{GL}(4,R)$ group-topology, covariance
  and curved-space spinors},''
  \href{http://dx.doi.org/10.1142/S0217751X87000867}{{\em Int. J. Mod. Phys. A}
  {\bf 2} (1987)  1655}.

\bibitem{Kopczynski:1988jq}
W.~Kopczynski, J.~D. McCrea, and F.~W. Hehl, ``{The Weyl group and its
  currents},'' \href{http://dx.doi.org/10.1016/0375-9601(88)90182-X}{{\em Phys.
  Lett. A} {\bf 128} (1988)  313--317}.

\bibitem{Stephani:2003tm}
H.~Stephani, D.~Kramer, M.~A. MacCallum, C.~Hoenselaers, and E.~Herlt,
  \href{http://dx.doi.org/10.1017/CBO9780511535185}{{\em {Exact solutions of
  Einstein's field equations}}}.
\newblock Cambridge Monographs on Mathematical Physics. Cambridge Univ. Press,
  Cambridge, 2003.

\bibitem{Ortin:2015hya}
T.~Ortin, \href{http://dx.doi.org/10.1017/CBO9781139019750}{{\em {Gravity and
  Strings}}}.
\newblock Cambridge Monographs on Mathematical Physics. Cambridge University
  Press, 2nd~ed., 7, 2015.

\bibitem{hartle1967variational}
J.~B. Hartle and D.~H. Sharp, ``Variational principle for the equilibrium of a
  relativistic, rotating star,'' {\em The Astrophysical Journal} {\bf 147}
  (1967)  317.

\bibitem{wu1976dirac}
T.~T. Wu and C.~N. Yang, ``{Dirac's monopole without strings: Classical
  Lagrangian theory},'' {\em Phys. Rev. D} {\bf 14} (1976) no.~2, 437.

\bibitem{Shnir_2005}
Y.~M. Shnir, \href{http://dx.doi.org/10.1007/3-540-29082-6}{{\em Magnetic
  Monopoles}}.
\newblock Springer Berlin Heidelberg, 2005.

\bibitem{Bahamonde:2021akc}
S.~Bahamonde and J.~Gigante~Valcarcel, ``{Observational constraints in
  metric-affine gravity},''
  \href{http://dx.doi.org/10.1140/epjc/s10052-021-09275-6}{{\em Eur. Phys. J.
  C} {\bf 81} (2021) no.~6, 495}, \href{http://arxiv.org/abs/2103.12036}{{\tt
  arXiv:2103.12036 [gr-qc]}}.

\bibitem{Newman:1965tw}
E.~T. Newman and A.~I. Janis, ``{Note on the Kerr spinning-particle metric},''
  \href{http://dx.doi.org/10.1063/1.1704350}{{\em J. Math. Phys.} {\bf 6}
  (1965)  915--917}.

\bibitem{Erbin:2014aya}
H.~Erbin, ``{Janis\textendash{}Newman algorithm: simplifications and gauge
  field transformation},''
  \href{http://dx.doi.org/10.1007/s10714-015-1860-1}{{\em Gen. Rel. Grav.} {\bf
  47} (2015)  19}, \href{http://arxiv.org/abs/1410.2602}{{\tt arXiv:1410.2602
  [gr-qc]}}.

\bibitem{Griffiths:2009dfa}
J.~B. Griffiths and J.~Podolský,
  \href{http://dx.doi.org/10.1017/CBO9780511635397}{{\em {Exact Space-Times in
  Einstein's General Relativity}}}.
\newblock Cambridge Monographs on Mathematical Physics. Cambridge University
  Press, Cambridge, 2009.

\bibitem{Podolsky:2021zwr}
J.~Podolský and A.~Vrátný, ``{New improved form of black holes of type D},''
  \href{http://dx.doi.org/10.1103/PhysRevD.104.084078}{{\em Phys. Rev. D} {\bf
  104} (2021) no.~8, 084078}, \href{http://arxiv.org/abs/2108.02239}{{\tt
  arXiv:2108.02239 [gr-qc]}}.

\end{thebibliography}\endgroup

\end{document}